\documentclass[%
reprint,
amsmath,amssymb,
aps,
twocolumn,
pra,
]{revtex4-2}

\usepackage{graphicx}
\usepackage{color}
\usepackage{dcolumn}
\usepackage{bm}
\usepackage{hyperref}

\usepackage{amsmath}
\usepackage{amssymb}
\usepackage{mathtools}
\usepackage{orcidlink}
\usepackage{soul} 
\usepackage{xcolor} 

\usepackage{tikz}
\usetikzlibrary{quantikz2}

\usepackage{cleveref}
\crefname{equation}{Eq.}{Eqs.}
\Crefname{equation}{Equation}{Equations}
\crefname{figure}{Fig.}{Figs.}
\Crefname{figure}{Figure}{Figures}
\crefname{table}{Table}{Tables}
\crefname{section}{Sec.}{Secs.}
\Crefname{section}{Section}{Sections}
\crefname{subsection}{Sec.}{Secs.}
\Crefname{subsection}{Section}{Sections}
\crefname{subsubsection}{Sec.}{Secs.}
\Crefname{subsubsection}{Section}{Sections}
\crefname{appendix}{Appendix}{Appendixes}
\Crefname{appendix}{Appendix}{Appendixes}
\crefalias{section}{appendix}

\newcommand{\qft}{\mathrm{QFT}}
\newcommand{\dt}{\Delta t}
\newcommand{\pdv}[3][]{\frac{\partial^{#1} #2}{\partial #3^{#1}}}

\begin{document}

\title{
	First-quantized quantum simulation with \\ non-local potentials by matrix-product-state encoding
}

\author{Kazuki Tsuoka\orcidlink{0009-0002-1892-1791}}
\email{kazukitsuoka@g.ecc.u-tokyo.ac.jp}

\affiliation{
	Department of Chemistry,
	School of Science,
	The University of Tokyo,
	Tokyo 113-0033,
	Japan
}

\author{Taichi Kosugi\orcidlink{0000-0003-3379-3361}}
\affiliation{
	Quemix Inc.,
	Taiyo Life Nihombashi Building,
	2-11-2,
	Nihombashi Chuo-ku,
	Tokyo 103-0027,
	Japan
}

\affiliation{
	Department of Physics,
	The University of Tokyo,
	Tokyo 113-0033,
	Japan
}

\author{Masari Watanabe\orcidlink{0000-0003-3186-535X}}
\affiliation{
	Quemix Inc.,
	Taiyo Life Nihombashi Building,
	2-11-2,
	Nihombashi Chuo-ku,
	Tokyo 103-0027,
	Japan
}

\affiliation{
	Department of Physics,
	The University of Tokyo,
	Tokyo 113-0033,
	Japan
}

\author{Hirofumi Nishi\orcidlink{0000-0001-5155-6605}}
\affiliation{
	Quemix Inc.,
	Taiyo Life Nihombashi Building,
	2-11-2,
	Nihombashi Chuo-ku,
	Tokyo 103-0027,
	Japan
}
\affiliation{
	Department of Physics,
	The University of Tokyo,
	Tokyo 113-0033,
	Japan
}
\author{Yu-ichiro Matsushita\orcidlink{0000-0002-9254-5918}}
\affiliation{
	Quemix Inc.,
	Taiyo Life Nihombashi Building,
	2-11-2,
	Nihombashi Chuo-ku,
	Tokyo 103-0027,
	Japan
}

\affiliation{Quantum Materials and Applications Research Center,
	National Institutes for Quantum Science and Technology (QST),
	2-12-1 Ookayama, Meguro-ku, Tokyo 152-8550, Japan
}

\affiliation{
	Department of Physics,
	The University of Tokyo,
	Tokyo 113-0033,
	Japan
}

\date{\today}

\begin{abstract}
Quantum simulations based on first-quantized approaches can offer lower space and gate complexities than second-quantized approaches.
The gate requirements for these simulations can be further reduced by employing pseudopotentials, a technique widely used in classical quantum chemistry calculations to decrease the number of simulated electrons.
However, implementing pseudopotentials in quantum circuits remains challenging due to their non-local nature.
In this study, we proposed a method to efficiently implement non-local potentials using matrix product state (MPS) encoding.
MPS encoding allows us to efficiently implement a Trotter step operator with a circuit depth of \(O(\deg(V_{\mathrm{loc}}) n^{\deg(V_{\mathrm{loc}})} + 8Lnn_{\mathrm{atom}}\chi^2 + 2n^2)\), where \(\deg(V_{\mathrm{loc}})\) is the degree of the polynomial local potential, \(n_{\mathrm{atom}}\) is the number of atoms in the system, \(L\) is the maximum number of projector functions across all atoms, and \(n\) is the number of qubits used to represent the binary-encoded wave function in one-dimensional space.
We demonstrated the effectiveness of this approach by applying it to a model of the ionization of a one-dimensional hydrogen atom driven by an ultrashort laser pulse.
Our results showed that MPS encoding significantly reduces the required circuit depth compared to a direct unitary decomposition, providing a pathway for practical first-quantized simulations.
\end{abstract}

\maketitle


\section{\label{sec:intro}Introduction}

The application of quantum computers to quantum chemistry problems has been extensively studied, including ground-state energy calculations~\cite{Aspuru-Guzik2005-qh}, geometry optimization~\cite{Kosugi2023-fu}, quasiparticle spectra~\cite{Kosugi2020-ou}, and linear-response functions~\cite{Kosugi2020-dh}. Simulation of the quantum dynamics of chemical systems is also one of the most promising applications of quantum computers~\cite{Miessen2023-cm}.

Quantum chemistry calculations on quantum computers, especially those that utilize variational methods, commonly use second-quantized Hamiltonians~\cite{Peruzzo2014-vq,McClean2016-mm,Cerezo2021-hs}.
This is because, for systems with a small number of particles in minimal basis sets, quantum chemistry calculations using second-quantized Hamiltonians can be performed with fewer qubits or gates, making them suitable for experimental demonstrations on near-term quantum computers~\cite{McArdle2020-jh}.
However, in the case of simulating the dynamics of chemical systems using a fault-tolerant quantum computer, Su et al.~\cite{Su2021-ev} found that first-quantized approaches were more efficient than state-of-the-art second-quantized approaches both in terms of the number of qubits and the number of gates required for simulation, and various studies have been conducted to utilize this advantage for quantum chemistry simulations using a real-space grid~\cite{Wiesner1996-gx,Zalka1998-oa,Benenti2008-xa,Kassal2008-ft,Kivlichan2017-gi,Ollitrault2020-ke}, plane-wave basis~\cite{Su2021-ev,Shokrian-Zini2023-tp,Berry2024-ez}, and arbitrary basis~\cite{Georges2025-ys}.

However, it can be difficult to simulate systems with a large number of electrons using first-quantized approaches, as the number of qubits required to represent the wave function increases linearly with the number of electrons~\cite{Su2021-ev,Kosugi2023-fu,Wiesner1996-gx,Zalka1998-oa,Kassal2008-ft,Kosugi2022-up,Cody-Jones2012-ej,Chan2023-ek}.
This problem can be mitigated by using a pseudopotential, which reduces the number of electrons to be simulated by replacing the effect of core electrons with an effective potential acting on valence electrons, and is widely used in classical quantum chemistry calculations~\cite{Kleinman1982-yz,Kresse1999-gh, Hamann1979-wm,Austin1962-cg,Dal-Corso2014-ij}.
One of the challenges in implementing pseudopotentials in quantum circuits is the implementation of arbitrary non-local potentials.
Although efficient implementation methods for the local Coulomb potential have been proposed using reflection symmetry~\cite{Huang2024-pt} and polynomial interpolation~\cite{Huang2025-hm}, applying these methods to non-local potentials, which are not diagonal in either the position or momentum basis, is not straightforward.
Several studies have proposed methods to implement pseudopotentials in first-quantized simulations~\cite{Shokrian-Zini2023-tp,Berry2024-ez,Georges2025-ys,da-Jornada2025-wi}.
Zini et al.~\cite{Shokrian-Zini2023-tp} made use of quantum read-only memory (QROM)~\cite{Babbush2018-no} to implement Goedecker-Teter-Hutter pseudopotentials~\cite{Goedecker1996-ze,Hartwigsen1998-vo}, but struggled with significant energy errors and high resource overheads. Subsequently, Berry et al.~\cite{Berry2024-ez} improved upon this approach by utilizing quantum arithmetic circuits to take advantage of their functional form, resulting in a more efficient and accurate implementation of the pseudopotentials.
Extending Berry et al.'s method for implementing pseudopotentials, da Jornada et al.~\cite{da-Jornada2025-wi} proposed a method to run non-Born-Oppenheimer dynamics simulations on a quantum computer. Their approach combines the nucleus and core electrons into a single particle, which they called a ``pseudoion,'' and implements the pseudopotential between the pseudoion and valence electrons based on Berry et al.'s method.
Georges et al.~\cite{Georges2025-ys} proposed a rather different approach to implement pseudopotentials in first-quantized simulations. In their approach, the Hamiltonian matrix elements are precalculated on a classical computer, converted into a linear combination of unitaries, and loaded into a quantum circuit using advanced QROM~\cite{Berry2019-os,Babbush2018-no,Low2024-nu}, enabling the use of modern pseudopotentials that cannot be expressed in a simple analytical form with any basis set.

In this study, we proposed a method to implement a non-local potential using matrix product state (MPS) encoding~\cite{Schon2005-dx,Schon2007-li,Lin2021-nt,Malz2024-cw,Ran2020-ud,Smith2024-pb} and investigated this method using a model of the ionization of a one-dimensional hydrogen atom by ultrashort laser pulses.
MPS is a class of tensor network states that is effective for representing quantum states with limited entanglement~\cite{Schollwock2011-kl} and is widely used in classical simulations of quantum many-body systems~\cite{Ostlund1995-jf, Klumper1992-hc, Fannes1992-sx, Vidal2003-vj}.
By truncating the bond dimension of the MPS, we can approximate the non-local potential with a low-rank MPS, thereby reducing the circuit depth required to implement it.
Various methods have been proposed to encode MPS into quantum circuits to reduce the circuit depth~\cite{Schon2005-dx,Schon2007-li,Lin2021-nt,Malz2024-cw,Ran2020-ud,Smith2024-pb}, and we adopted a simple method~\cite{Lin2021-nt} that can exactly encode MPS into quantum circuits with a depth of \(O(n\chi^2)\), where \(n\) is the number of qubits and \(\chi\) is the maximum bond dimension of the MPS.

\section{\label{sec:background}Methods}

\subsection{\label{sec:realspace}Quantum simulation of first-quantized real-space Hamiltonians}
To introduce this method, we first consider a one-dimensional single-electron system and explain how the dynamics of this system can be simulated using a quantum computer.
To digitally represent the wave function of an electron in one-dimensional space, we need to discretize the position space into \(N\) grid points. The grid points are given by
\begin{align}
	x_j = x_0 + j \Delta x \quad (j = 0, 1, \ldots, N - 1),
\end{align}
where \(\Delta x\) is the spacing between adjacent grid points.
The wave function \(\ket{\psi}\) can be represented on these grid points as
\begin{equation}
	\ket{\psi} = \sqrt{\Delta x} \sum_{j=0}^{N-1}\psi(x_j)\ket{j},
\end{equation}
where \(\psi(x_j)\) is the value of the wave function at the \(j\)-th grid point, and \(\ket{j}\) is the computational basis state corresponding to the \(j\)-th grid point.
We introduced the factor $\sqrt{\Delta x}$ to obey the normalization condition
\begin{equation}
	{\braket{\psi}{\psi}} = \sum_{j=0}^{N-1}\Delta x\,|\psi(x_j)|^2=1.
\end{equation}
We assume that \(N\) is a power of 2, i.e., \(N = 2^n\), where \(n\) is the number of qubits used to represent the wave function.
In this case, the computational basis state can be expressed as a tensor product of single-qubit states:
\begin{equation}
	\ket{j}=\ket{j_0}\otimes\ket{j_1}\otimes\cdots\otimes\ket{j_{n-1}},
\end{equation}
where \(j = j_0 2^0 + j_1 2^1 + \cdots + j_{n-1} 2^{n-1}\) and \(j_k \in \{0, 1\}\) for \(k = 0, 1, \ldots, n-1\). This representation allows us to efficiently encode the discrete wave function of the electron with \(2^n\) grid points in a quantum computer using \(n\) qubits.
The Hamiltonian of an electron in one-dimensional space in atomic units is given by
\begin{align}
	H_{\mathrm{1D}} = T_{\mathrm{kin}} + V,
\end{align}
where
\begin{align}
	T_{\mathrm{kin}} = -\frac{1}{2} \pdv[2]{}{x}, \quad V = V(x),
\end{align}
are the kinetic energy operator and the potential energy operator, respectively.
By first-order Trotterization, the time-evolution operator for a time step of \(\dt\) can be approximated as
\begin{equation}
	\ket{\psi (t)} = e^{-iH_{\mathrm{1D}}t}\ket{\psi (0)} \approx \left(e^{-iT_{\mathrm{kin}}\Delta t} e^{-iV\Delta t}\right)^{t/\Delta t}\ket{\psi (0)},
\end{equation}
with a leading error of the order of \(O(\dt^2)\) per step.
To simplify the implementation of the time-evolution operator using gates, we can diagonalize the kinetic energy operator \(T_{\mathrm{kin}}\) by transforming it into the momentum basis using a discrete Fourier transform.
Let $\mathcal F$ denote the unitary discrete Fourier transform matrix that maps the momentum-grid basis to the position-grid basis, with matrix elements
\begin{equation}
	\mathcal F_{jk}=\frac{1}{2^{n/2}}\exp\left(\frac{2\pi i jk}{2^n}\right),
\end{equation}
so that \(\tilde{T}_{\mathrm{kin}} = \mathcal{F}^\dagger T_{\mathrm{kin}} \mathcal{F}\) is diagonal in the momentum-grid basis, leading to
\begin{equation}
	e^{-iT_{\mathrm{kin}}\Delta t} = \mathcal{F} e^{-i\tilde{T}_{\mathrm{kin}}\Delta t} \mathcal{F}^\dagger.
\end{equation}
Therefore, we can repeatedly apply the Trotter step operator to simulate the time evolution of the wave function on a quantum computer:
\begin{equation}
	\ket{\psi(t+\Delta t)} \approx \mathcal{F} e^{-i\tilde{T}_{\mathrm{kin}}\Delta t} \mathcal{F}^\dagger e^{-iV\Delta t} \ket{\psi (t)}.
\end{equation}
The simulation of a multi-electron system can be performed in a similar manner by assigning \(n\) qubits to each electron and using the tensor product of the single-electron Hilbert spaces to represent the spatial part of the multi-electron wave function~\cite{Kosugi2023-fu,Wiesner1996-gx,Zalka1998-oa,Kassal2008-ft,Kosugi2022-up,Cody-Jones2012-ej,Chan2023-ek}. If the system consists of \(\eta\) electrons, the total number of grid points in the position space is \(2^{n\eta}\), making it difficult to simulate the dynamics of the system on a classical computer. However, a quantum computer can efficiently represent the wave function of the multi-electron system using \(n\eta\) qubits, showing an exponential advantage over classical computers in terms of memory usage.

\subsection{\label{sec:mpscircuit}MPS and encoding circuits}
In an MPS, a many-body quantum state is expressed as a product of matrices, one for each site. An MPS representation of a quantum state \(\ket{\psi}\) can be written as
\begin{multline}\label{eq:mps_def}
	\ket{\psi} = \sum_{j_0, \ldots, j_{n-1}=0}^1 \sum_{w_1=1}^{\chi_1} \cdots \sum_{w_{n-1}=1}^{\chi_{n-1}} \\
	A_{1 w_1}^{[0] j_0} A_{w_1 w_2}^{[1] j_1} \cdots A_{w_{n-1} 1}^{[n-1] j_{n-1}}\left(\ket{j_0}\otimes\ket{j_1}\otimes\cdots\otimes\ket{j_{n-1}}\right),
\end{multline}
where \(w_k\) is the bond index that connects the \((k-1)\)-th and \(k\)-th sites, which runs from 1 to \(\chi_k\), the bond dimension of the MPS, and \(A^{[k]}\) is a complex-valued tensor with dimensions of \(\chi_{k} \times 2 \times \chi_{k+1}\). We define the maximum bond dimension of the MPS as \(\chi = \max_{k \in \{1, \ldots, n-1\}} \chi_k\).

To exactly encode an MPS into a quantum circuit, we adopted a method from~\cite{Lin2021-nt}. When an MPS is in the right-canonical form, i.e., matrix elements \(A_{w_{k} w_{k+1}}^{[k] j_k}\) satisfy
\begin{equation}
	\sum_{j_k=0}^1 \sum_{w_{k+1}=1}^{\chi_{k+1}} A_{w_{k} w_{k+1}}^{[k] j_k} A_{w_{k}' w_{k+1}}^{[k] j_k *} = \delta_{w_{k} w_{k}'},
\end{equation}
meaning that \(A^{[k]}\) works as an isometry that maps \(\ket{w_k} \in \mathbb{C}^{\chi_k}\) to \(\ket{j_k}\otimes\ket{w_{k+1}} \in \mathbb{C}^{2}\times\mathbb{C}^{\chi_{k+1}}\). By completing this isometry to a unitary operator \(U^{[k]}\) that maps \(\ket{w_k}\otimes\ket{0}^{\otimes \log_2(2\chi_{k+1}/\chi_k)}\) to \(\ket{j_k}\otimes\ket{w_{k+1}}\) as
\begin{multline}\label{eq:mps_site_as_unitary}
	A_{w_{k} w_{k+1}}^{[k] j_k} \\
	= \left(\bra{j_k}\otimes\bra{w_{k+1}}\right)U^{[k]}\left(\ket{w_k}\otimes\ket{0}^{\otimes \log_2(2\chi_{k+1}/\chi_k)}\right),
\end{multline}
and substituting this into \cref{eq:mps_def}, we can sequentially apply these unitaries to the initialized multi-qubit state \(\ket{0}^{\otimes n}\) to prepare \(\ket{\psi}\) as
\begin{equation}\label{eq:mps_prep_unitary}
	\ket{\psi} = U^{[n-1]} U^{[n-2]} \cdots U^{[0]} \ket{0}^{\otimes n}.
\end{equation}
Each unitary \(U^{[k]}\) acts on at most \(\log_2 \chi + 1\) sites.
Further decomposing an \(m\)-qubit unitary into single-qubit gates and two-qubit gates requires \(O(4^m)\) gates to implement~\cite{Krol2024-ys}, so the depth of the circuit required to encode an \(n\)-site MPS with maximum bond dimension of \(\chi\) is \(O(n4^{\log_2 \chi + 1}) = O(4n\chi^2)\).

\subsection{\label{sec:generalimpl}MPS representation of non-local potentials}
A norm-conserving pseudopotential $V$ is commonly decomposed into a local part $V_{\mathrm{loc}}$ and a non-local component $V_{\mathrm{nonl}}$, the latter being expressed as a finite sum of separable projectors:
\begin{equation}
	V = V_{\mathrm{loc}} + V_{\mathrm{nonl}}.
\end{equation}
Let $n_{\mathrm{atom}}$ be the number of atoms in the system.
The non-local part is generally written in the Kleinman-Bylander form~\cite{Kleinman1982-yz} as:
\begin{equation}
	V_{\mathrm{nonl}}=
	\sum_{\nu=0}^{n_{\mathrm{atom}}-1}
	\sum_{s\in\mathcal S_\nu}
	d_{\nu s}\ket{\phi_{\nu s}}\bra{\phi_{\nu s}},
\end{equation}
where $\mathcal{S}_\nu$ is the set of channels associated with atom $\nu$, $d_{\nu s}\in\mathbb{R}$ are coefficients, and \(\ket{\phi_{\nu s}}\) is a localized one-electron projector function centered at the position \(\bm{R}_\nu\) of atom \(\nu\).
Here, we assume that the projector functions at each atom are orthonormalized and those at different atoms do not overlap:
\begin{align}
	\langle \phi_{\nu s} | \phi_{\nu' s'} \rangle
	=
	\delta_{\nu \nu'}
	\delta_{s s'}.
\end{align}
Thus, the non-local potential propagator $U_{\mathrm{nonl}} = \exp(-i V_{\mathrm{nonl}} \Delta t)$ can be expressed as:
\begin{align}
	U_{\mathrm{nonl}} & = \exp(-i\Delta t \sum_{\nu=0}^{n_{\mathrm{atom}}-1} \sum_{s\in\mathcal S_\nu} d_{\nu s} | \phi_{\nu s} \rangle \langle \phi_{\nu s} |) \nonumber                      \\
	                  & = \prod_{\nu=0}^{n_{\mathrm{atom}}-1} \prod_{s\in\mathcal S_\nu} \exp(-i d_{\nu s} \Delta t | \phi_{\nu s} \rangle \langle \phi_{\nu s} |).\label{eq:U_nonl_decomposed}
\end{align}

\begin{figure*}[t]
	\centering
	\scalebox{0.91}{
		\begin{quantikz}[column sep=0.16cm, row sep=0.2cm]
			\lstick[4]{\(\ket{\psi(t)}\)} & \gate[4]{e^{-iV_{\mathrm{loc}}\Delta t}} & \qw\gategroup[4,steps=11,style={dashed,inner sep=0pt}]{Equal to \(U_{\mathrm{nonl}}\), loop over all atoms \(s \in \mathcal{S}_\nu\) and \(\nu\) from 0 to \(n_{\mathrm{atom}} - 1\)} & \qw                              & \qw                              & \gate[2]{U_{\nu s}^{[0]\dagger}} & \gate{X} & \ctrl{1}                      & \gate{X} & \gate[2]{U_{\nu s}^{[0]}} & \qw                       & \qw                       & \qw                    & \gate[4]{\qft^\dagger} & \gate[4]{e^{-i\tilde{T}_{\mathrm{kin}}\Delta t}} & \gate[4]{\qft} & \qw \\
			                            & \qw                                      & \qw                                                                                                                                                                                   & \qw                              & \gate[3]{U_{\nu s}^{[1]\dagger}} & \qw                              & \gate{X} & \ctrl{1}                      & \gate{X} & \qw                       & \gate[3]{U_{\nu s}^{[1]}} & \qw                       & \qw                    & \qw                    & \qw                                              & \qw            & \qw \\
			                            & \qw                                      & \qw                                                                                                                                                                                   & \gate[2]{U_{\nu s}^{[2]\dagger}} & \qw                              & \qw                              & \gate{X} & \ctrl{1}                      & \gate{X} & \qw                       & \qw                       & \gate[2]{U_{\nu s}^{[2]}} & \qw                    & \qw                    & \qw                                              & \qw            & \qw \\
			                            & \qw                                      & \gate{U_{\nu s}^{[3]\dagger}}                                                                                                                                                         & \qw                              & \qw                              & \qw                              & \gate{X} & \gate{P(-d_{\nu s} \Delta t)} & \gate{X} & \qw                       & \qw                       & \qw                       & \gate{U_{\nu s}^{[3]}} & \qw                    & \qw                                              & \qw            & \qw
		\end{quantikz}
	}
	\caption{An example 4-qubit quantum circuit diagram implementing a single Trotter step operator using a full-rank MPS representation of the non-local potential. We omit the translation operator \(T(\bm{R}_\nu)\). \(P(\phi) = \ket{0}\bra{0}+e^{i\phi}\ket{1}\bra{1}\) is a phase gate, and ``QFT'' represents the quantum Fourier transform circuit, which implements the discrete Fourier transform \(\mathcal{F}\).}\label{fig:non-local_potential_circuit}
\end{figure*}
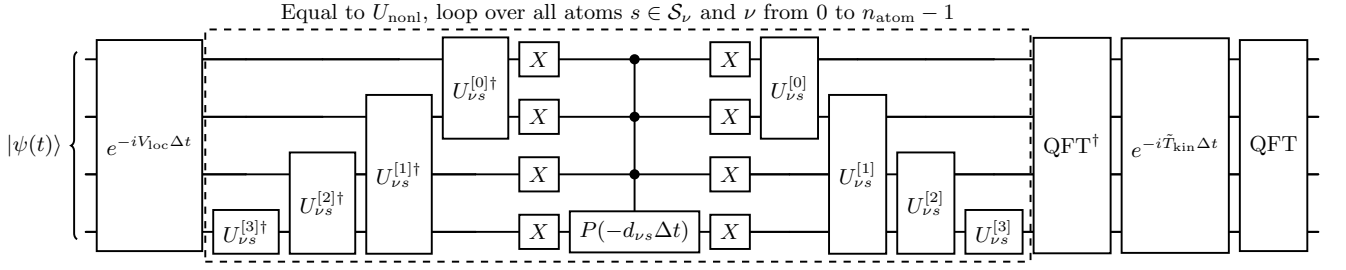

To implement the non-local potential operator on a quantum computer, we represent the projector functions $| \phi_{\nu s} \rangle$ using MPS.
We first represent each origin-centered projector function \(\ket{\phi_{\nu s,\mathrm{orig}}}\) by an MPS whose maximum bond dimension is denoted by \(\chi_{\nu s}\), and encode it into a quantum circuit using the method described in \cref{sec:mpscircuit}.
Let $U_{\nu s}$ be a state-preparation unitary satisfying
\begin{equation}
	\ket{\phi_{\nu s,\mathrm{orig}}} = U_{\nu s}\ket{\mathbf{0}} = U_{\nu s}^{[n-1]} U_{\nu s}^{[n-2]} \cdots U_{\nu s}^{[0]} \ket{\mathbf{0}},
\end{equation}
where \(\ket{\mathbf{0}}\equiv\ket{0}^{\otimes n}\). Here, \(U_{\nu s}^{[k]}\) is a unitary as in \cref{eq:mps_prep_unitary}.
To avoid creating multiple matrix product states for the same projector function located at different atom positions, we introduce the translation operator $T(\bm{R}_\nu)$ that translates the state localized at the origin to the position $\bm{R}_\nu$ of atom $\nu$ as
\begin{align}
	| \phi_{\nu s} \rangle = T(\bm{R}_\nu) U_{\nu s} \ket{\mathbf{0}}.
\end{align}
The translation operator can be implemented using the Fourier transform as follows:
\begin{align}
	T(\bm{R}_\nu) = \mathcal{F} U_{\mathrm{ph}}(\bm{R}_\nu) \mathcal{F}^\dagger,
\end{align}
where $U_{\mathrm{ph}}(\bm{R}_\nu)$ is a tensor product of single-qubit phase gates~\cite{Kosugi2024-la}.
By substituting these expressions into \cref{eq:U_nonl_decomposed}, we can express $U_{\mathrm{nonl}}$ as:
\begin{widetext}
	\begin{equation}
		U_{\mathrm{nonl}}
		=
		\prod_{\nu = 0}^{n_{\mathrm{atom}} - 1}
		T (\bm{R}_\nu)
		\left(
		\prod_{s\in\mathcal S_\nu}
		U_{\nu s}
		U_0 (d_{\nu s} \Delta t)
		U_{\nu s}^\dagger
		\right)
		T (\bm{R}_\nu)^\dagger
		=
		\mathcal{F}
		\left(
		\prod_{\nu = 0}^{n_{\mathrm{atom}} - 1}
		U_{\mathrm{ph}} (\bm{R}_\nu)
		\mathcal{F}^\dagger
		W_\nu
		\mathcal{F}
		U_{\mathrm{ph}} (\bm{R}_\nu)^\dagger
		\right)
		\mathcal{F}^\dagger,
	\end{equation}
\end{widetext}
where
\begin{align}\label{eq:U_0c}
	U_0(c) = \exp(-i c \ket{\mathbf{0}}\bra{\mathbf{0}}),
\end{align}
and
\begin{align}\label{eq:W_nu}
	W_\nu
	\equiv
	\prod_{s\in\mathcal S_\nu}
	U_{\nu s}
	U_0 (d_{\nu s} \Delta t)
	U_{\nu s}^\dagger.
\end{align}
Thus, the Trotter step operator \(e^{-iH\Delta t}\) can be approximated via the first-order Trotterization formula as
\begin{equation}\label{eq:trotterstep}
	e^{-iH\Delta t} \approx \mathcal{F} e^{-i\tilde{T}_{\mathrm{kin}}\Delta t} \mathcal{F}^\dagger U_{\mathrm{nonl}} e^{-iV_{\mathrm{loc}}\Delta t}.
\end{equation}
An example of a quantum circuit implementing a single Trotter step operator using an MPS representation of the non-local potential is shown in \cref{fig:non-local_potential_circuit}.
\(U_0(c)\) can be implemented using a single multi-controlled phase gate, which has an \(O(n)\) depth decomposition algorithm~\cite{da-Silva2022-fw}.
Thus, the depth of the circuit required to implement \(W_\nu\) is \(O(L(2\cdot 4n\chi^2 + n)) = O(8Ln\chi^2)\), where \(L = \max_\nu |\mathcal{S}_\nu|\) is the maximum number of projector functions \(|\phi_{\nu s}\rangle\) across all atoms \(\nu\) and \(\chi = \max_{\nu,s} \chi_{\nu s}\) denotes the maximum bond dimension across all MPS representations of the projector functions \(|\phi_{\nu s}\rangle\).
\(U_{\mathrm{ph}} (\bm{R}_\nu)\) is merely a tensor product of single-qubit phase gates; therefore, it can be implemented with \(O(1)\) depth. The QFT requires \(O(n)\) depth~\cite{Fowler2004-by}, so the total depth of the circuit required to implement \(U_{\mathrm{nonl}}\) is \(O(n_{\mathrm{atom}}(8Ln\chi^2 + 2n + 2)) = O(8Lnn_{\mathrm{atom}}\chi^2)\).
It is known that the polynomial phase gate \(\sum_{j=0}^{2^n-1} e^{-i f(x_j)}|j\rangle\langle j|\), where \(f(x) = a_0 + a_1 x + \cdots + a_M x^M\) is a degree-\(M\) real polynomial function, can be efficiently implemented with \(O(Mn^M)\) depth~\cite{Benenti2008-xa,Ollitrault2020-ke,Kosugi2023-fu,Huang2025-hm}, so \(e^{-i\tilde{T}_{\mathrm{kin}}\Delta t} = \sum_{j=0}^{2^n-1} e^{-i{p_j}^2\Delta t/2} |j\rangle\langle j|\), where \(p_j\) is the momentum value assigned to the grid index \(j\) (defined in \cref{sec:split_operator_results}), can be implemented with \(O(2n^2)\) depth.
Assuming that the local part of the pseudopotential \(V_{\mathrm{loc}}\) is a real polynomial function of degree \(\deg(V_{\mathrm{loc}})\), \(e^{-iV_{\mathrm{loc}}\Delta t}\) can be implemented with \(O(\deg(V_{\mathrm{loc}}) n^{\deg(V_{\mathrm{loc}})})\) depth.
Therefore, in total, each Trotter step requires \(O(\deg(V_{\mathrm{loc}}) n^{\deg(V_{\mathrm{loc}})} + 8Lnn_{\mathrm{atom}}\chi^2 + n + 2n^2 + n) = O(\deg(V_{\mathrm{loc}}) n^{\deg(V_{\mathrm{loc}})} + 8Lnn_{\mathrm{atom}}\chi^2 + 2n^2)\) depth.

\subsection{\label{sec:model}Ionization of a one-dimensional hydrogen atom by an ultrashort laser pulse}

To benchmark the proposed method for constructing non-local potential propagators, we considered the ionization of a one-dimensional hydrogen atom by an ultrashort laser pulse as a model system.
The time-dependent Schr\"odinger equation (TDSE) in the momentum representation governing the dynamics of a three-dimensional hydrogen atom is given by
\begin{multline}\label{eq:3d_tdse}
	\left[i \frac{\partial}{\partial t}-\frac{p^2}{2}- A(t)(\bm{e} \cdot \bm{p})\right] \Phi_{\mathrm{3D}}(\bm{p}, t) \\
	-\int \frac{d^3 \bm{p}^{\prime}}{(2 \pi)^3} V\left(\bm{p}, \bm{p}^{\prime}\right) \Phi_{\mathrm{3D}}\left(\bm{p}^{\prime}, t\right)=0.
\end{multline}
In \cref{eq:3d_tdse}, \(\bm{e}\) is the polarization unit vector, and \(A(t)\) is the vector potential of the external electric field, defined as
\begin{equation}
	A(t) =
	\begin{dcases}
		A_0 \sin(\omega t) \sin^2\!\left(\frac{\pi t}{\tau_{\mathrm{pulse}}}\right) & (0 \leq t \leq \tau_{\mathrm{pulse}}), \\
		0                                                                          & (\text{otherwise}).
	\end{dcases}
\end{equation}
Here, \(A_0 = \omega^{-1}\sqrt{I_{\mathrm{peak}}/I_0}\), where \(I_{\mathrm{peak}}\) is the peak
intensity of the laser pulse, \(I_0 = 3.5\times10^{16}\,\mathrm{W/cm^2}\) is the atomic unit
of intensity, \(\omega\) is the laser frequency, and \(\tau_{\mathrm{pulse}} = 2\pi n_{\mathrm{cycle}}/\omega\) is the pulse duration with \(n_{\mathrm{cycle}}\) being the number of cycles.
The Coulomb kernel \(V(\bm{p}, \bm{p}^{\prime})\) in momentum space is non-local. It is obtained as the Fourier transform of the electron-nuclear Coulomb potential,
\begin{equation}
	V(\bm{p}, \bm{p}^{\prime}) = - \frac{4 \pi}{|\bm{p} - \bm{p}^\prime|^2}.
\end{equation}
This kernel can be replaced by a
separable potential~\cite{Giraud2006-bg,Tetchou-Nganso2007-br,Tetchou-Nganso2011-ie,Tetchou-Nganso2013-zs}:
\begin{equation}
	V\left(\bm{p}, \bm{p}^{\prime}\right) \to - \frac{16 \pi}{(p^2 + 1)(p^{\prime 2} + 1)} = - v_{\mathrm{1s}}(p) v_{\mathrm{1s}}^*(p^\prime),
\end{equation}
where \(v_{\mathrm{1s}}(p) = 4 \sqrt{\pi}/(p^2 + 1)\), with $p = |\bm{p}|$ and $p' = |\bm{p}'|$.
The separable potential is constructed so as to reproduce exactly the momentum-space \(1s\) eigenstate obtained with the original Coulomb kernel,
\begin{equation}
	\psi_{\mathrm{3D,1s}}(p)=\frac{8 \sqrt{\pi}}{\left(p^2+1\right)^2},
\end{equation}
which we take as the initial state, \(\Phi_{\mathrm{3D}}(\bm{p}, 0) = \psi_{\mathrm{3D,1s}}(p)\).

For simplicity, we further reduced the model in \cref{eq:3d_tdse} to a one-dimensional hydrogen atom as follows:
\begin{multline}\label{eq:1d_tdse}
	\left[i \frac{\partial}{\partial t}-\frac{p^2}{2}- A(t) p\right] \Phi_{\mathrm{1D}}(p, t) \\
	-\int \frac{dp^\prime}{2 \pi} V(p, p^\prime) \Phi_{\mathrm{1D}}\left(p^{\prime}, t\right)=0.
\end{multline}
The one-dimensional Coulomb potential, \(V(x)=-1/|x|\), is singular at the origin. We therefore replace it with the regularized soft Coulomb potential and define \(V(p, p')\) as the Fourier transform of
\begin{equation}
	V_{\mathrm{soft}}(x) = - \frac{1}{\sqrt{x^2 + \epsilon^2}},
\end{equation}
where \(\epsilon\) is a softening parameter, and we set \(\epsilon = 1\) in this study.
The field-free Hamiltonian of the one-dimensional hydrogen atom with the soft Coulomb potential is given by
\begin{equation}\label{eq:1D_soft_H}
	H_{\mathrm{1D,soft}} = -\frac{1}{2} \pdv[2]{}{x} + V_{\mathrm{soft}}(x).
\end{equation}
The ground state and the first excited state of this Hamiltonian can be approximated by the following Gaussian-type functions~\cite{Grasselli2017-xx}:
\begin{align}
	\phi_0 (p) & = \left(\frac{4\pi}{\beta_0}\right)^{1/4} e^{-\frac{p^2}{2\beta_0}},         \\
	\phi_1 (p) & = -2ip \left( \frac{\pi}{\beta_1^3} \right)^{1/4} e^{-\frac{p^2}{2\beta_1}},
\end{align}
where \(\beta_0\) and \(\beta_1\) are variational parameters.
These parameters are determined by minimizing the expectation values \(E_0\) and \(E_1\) of the Hamiltonian, given by
\begin{align}
	E_0 & = \int dx \phi_0^*(x) H_{\mathrm{1D,soft}} \phi_0(x), \\
	E_1 & = \int dx \phi_1^*(x) H_{\mathrm{1D,soft}} \phi_1(x).
\end{align}
Here, \(\phi_0(x)\) and \(\phi_1(x)\) denote the coordinate-space counterparts of \(\phi_0(p)\) and \(\phi_1(p)\), respectively.

In the following simulation, we replaced \(V(p, p')\) with either a single-separable potential,
\begin{equation}\label{eq:single_separable}
	V(p, p^\prime) \to - v_{0}(p) v_{0}^*(p^\prime),
\end{equation}
where
\begin{equation}\label{eq:v0}
	v_0(p) = \frac{p^2 -2E_0}{ \sqrt{\beta_0 - 4E_0}} \phi_0(p),
\end{equation}
or a double-separable potential,
\begin{equation}\label{eq:double_separable}
	V(p, p^\prime) \to - v_{1}(p) v_{1}^*(p^\prime) - v_{2}(p) v_{2}^*(p^\prime),
\end{equation}
where
\begin{align}
	v_1(p) & =\alpha_{11}\left(E_0-\frac{1}{2} p^2\right) \phi_0(p)+\alpha_{12}\left(E_1-\frac{1}{2} p^2\right) \phi_1(p), \\
	v_2(p) & =\alpha_{21}\left(E_0-\frac{1}{2} p^2\right) \phi_0(p)+\alpha_{22}\left(E_1-\frac{1}{2} p^2\right) \phi_1(p).
\end{align}
The coefficients \(\alpha_{11}, \alpha_{12}, \alpha_{21}\), and \(\alpha_{22}\) are set to
\begin{equation}\label{eq:a_variables_def}
	\alpha_{11}=-\frac{1}{\sqrt{\gamma_{11}}}, \quad \alpha_{12}=0, \quad \alpha_{21}=0, \quad \alpha_{22}=\frac{1}{\sqrt{\gamma_{22}}}.
\end{equation}
The parameters \(\gamma_{11}\) and \(\gamma_{22}\) are calculated using \cref{eq:gamma11,eq:gamma22} (see \cref{app:derivation,app:derivation2} for details).
The two replacements are constructed so that, in the resulting Hamiltonian with separable potentials, \(\phi_0(p)\) alone becomes an exact eigenstate with eigenvalue \(E_0\) for the single-separable case, whereas \(\phi_0(p)\) and \(\phi_1(p)\) become exact eigenstates with eigenvalues \(E_0\) and \(E_1\), respectively, for the double-separable case.
As in the three-dimensional case, we take the exact ground state of the resulting Hamiltonian with separable potentials as the initial state, \(\Phi_{\mathrm{1D}}(p, 0) = \phi_0(p)\).

\section{\label{sec:results}Results and discussion}

\subsection{\label{sec:split_operator_results}Simulation of a one-dimensional hydrogen atom interacting with an ultrashort laser pulse using the split-operator method with a single-separable potential}
\begin{figure}
	\centering
	\includegraphics[width=\linewidth]{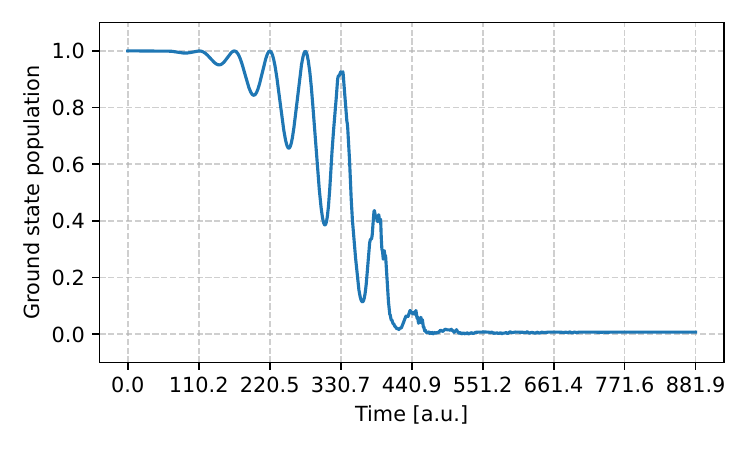}
	\caption{\label{fig:survival_probability}The temporal evolution of the ground-state population \(|\langle \phi_0 | \psi(t) \rangle|^2\) of a one-dimensional hydrogen atom interacting with an 8-cycle ultrashort laser pulse with peak intensity of \(I_{\mathrm{peak}} = 4.0 \times 10^{14} \, \mathrm{W/cm^2}\) and angular frequency of \(\omega = 0.057 \, \mathrm{a.u.}\), simulated by the split-operator method with the single-separable potential given in \cref{eq:single_separable}.}\label{fig:survival}
\end{figure}

\Cref{fig:survival} shows the time evolution of the one-dimensional hydrogen atom interacting with an ultrashort laser pulse. The laser pulse parameters were set to \(n_{\mathrm{cycle}} = 8\) optical cycles, a peak intensity of \(I_{\mathrm{peak}} = 4.0 \times 10^{14} \, \mathrm{W/cm^2}\), and an angular frequency of \(\omega = 0.057 \, \mathrm{a.u.}\) The simulation was performed using the first-order split-operator method with $2^{10} = 1024$ grid points and 10,000 time steps.
The grid points \(\{p_j\}\) were uniformly distributed in momentum space from \(-p_{\mathrm{max}} = -3 \, \mathrm{a.u.}\) to \(p_{\mathrm{max}} = 3 \, \mathrm{a.u.}\), given by
\begin{equation}
	p_j = -p_{\mathrm{max}} + j\Delta p, \qquad \Delta p = \frac{2p_{\mathrm{max}}}{2^n-1},
\end{equation}
for \(j = 0, 1, \ldots, 2^n-1\). The time step was set to \(\Delta t = \tau_{\mathrm{pulse}}/10000 \approx 0.088 \, \mathrm{a.u.}\)
In this case, the Hamiltonian governing the dynamics of the discretized wave function of the one-dimensional hydrogen atom is expressed as
\begin{equation}
	H_{\mathrm{1D,nonl}} = \tilde{T}_{\mathrm{kin}} + V_{\mathrm{nonl}} + V_{\mathrm{loc}}(t),
\end{equation}
where the kinetic energy operator \(\tilde{T}_{\mathrm{kin}}\), the non-local single-separable potential operator \(V_{\mathrm{nonl}}\), and the local potential operator \(V_{\mathrm{loc}}(t)\) describing the laser field interaction are respectively defined as:
\begin{align}
	\tilde{T}_{\mathrm{kin}} &= \sum_{j=0}^{2^n-1} \frac{p_j^2}{2} \ket{j}\bra{j}, \\
	V_{\mathrm{nonl}} &= -\ket{v_0}\bra{v_0}, \\
	V_{\mathrm{loc}}(t) &= \sum_{j=0}^{2^n-1} A(t)p_j \ket{j}\bra{j},
\end{align}
with \(\ket{v_0} = \sqrt{\Delta p/(2\pi)}\sum_{j=0}^{2^n-1} v_0(p_j) \ket{j}\).

We used the following first-order approximation for the time-evolution operator for a time step of \(\Delta t\):
\begin{equation}\label{eq:split_operator_1dh}
	e^{-iH_{\mathrm{1D,nonl}}\Delta t} \approx e^{-i\tilde{T}_{\mathrm{kin}}\Delta t} e^{-iV_{\mathrm{nonl}}\Delta t} e^{-iV_{\mathrm{loc}}(t)\Delta t}.
\end{equation}
Note that we did not use the Fourier transform because the potential is already in the momentum representation.

\subsection{\label{sec:circuit}Quantum circuit implementation of the time-evolution operator for a one-dimensional hydrogen atom using MPS representation of non-local separable potentials}
In this section, we describe the quantum circuit implementation of the time-evolution operator for the one-dimensional hydrogen atom model, given by \cref{eq:split_operator_1dh}.

\subsubsection{\label{sec:impl_v_loc}Implementation of \(e^{-iV_{\mathrm{loc}}(t)\Delta t}\)}
\(V_{\mathrm{loc}}(t)\) is diagonal in the momentum representation, acting on each basis state \(\ket{j}\) as
\begin{equation}
	V_{\mathrm{loc}}(t) \ket{j} = A(t)p_j \ket{j}.
\end{equation}
Let \(\tau_1 = A(t)\Delta t\). We can rewrite \(j\) in its binary representation as
\begin{align}
	j = \sum_{k=0}^{n-1} 2^{k} j_{k} = \sum_{k=0}^{n-1} 2^{k} \delta_{j_{k}, 1},
\end{align}
where \(\delta_{k, l}\) is the Kronecker delta function, which is 1 if \(k=l\) and 0 otherwise. Therefore, we have
\begin{equation}
	e^{-i \tau_1 \Delta p j} = \prod_{k=0}^{n-1} e^{-i2^k \tau_1 \Delta p \delta_{j_{k}, 1}}.
\end{equation}
The factor \(e^{-i2^k \tau_1 \Delta p \delta_{j_{k}, 1}}\) applies a \(2^k \tau_1 \Delta p\) phase shift to the state \(\ket{j}\) if \(j_{k} = 1\), which can be implemented by a \(P(-2^k \tau_1 \Delta p)\) gate acting on the \(k\)-th qubit.
Thus, the operator can be expanded as
\begin{equation}
    e^{-iV_{\mathrm{loc}}(t)\Delta t} = \sum_{j=0}^{2^n-1} e^{-i\tau_1 (-p_{\mathrm{max}} + j\Delta p)}\ket{j}\bra{j},
\end{equation}
which is equivalent to a tensor product of single-qubit phase gates up to a global phase:
\begin{equation}\label{eq:V_loc_exp_impl}
    e^{-iV_{\mathrm{loc}}(t)\Delta t} = e^{i\tau_1 p_{\mathrm{max}}} \bigotimes_{k=0}^{n-1} P(-2^k \tau_1 \Delta p).
\end{equation}

\subsubsection{\label{sec:impl_v_nonl}Implementation of \(e^{-iV_{\mathrm{nonl}}\Delta t}\)}
As described in \cref{sec:generalimpl}, the non-local potential operator \(V_{\mathrm{nonl}}\) can be implemented using the MPS representations of \(\ket{v_0}\), \(\ket{v_1}\) and \(\ket{v_2}\) as in \cref{eq:single_separable} or \cref{eq:double_separable}.
First, let us consider the single-separable potential case in \cref{eq:single_separable}.
Let \(U_{v_0}\) be a state-preparation unitary satisfying \(\ket{v_0}/\sqrt{N_{v_0}\Delta p/(2\pi)} = U_{v_0}\ket{\mathbf{0}}\), where \(N_{v_0} = \sum_{j=0}^{2^n-1} |v_0(p_j)|^2\) is the normalization factor of \(\ket{v_0}\).
Then, the operator \(e^{-iV_{\mathrm{nonl}}\Delta t} = e^{i\Delta t \ket{v_0}\bra{v_0}}\) can be implemented as
\begin{equation}
	e^{-iV_{\mathrm{nonl}}\Delta t} = U_{v_0} e^{i N_{v_0} \Delta t\Delta p/(2\pi) \ket{\mathbf{0}}\bra{\mathbf{0}}} U_{v_0}^\dagger.
\end{equation}
The central rotation \(e^{i N_{v_0} \Delta t\Delta p/(2\pi) \ket{\mathbf{0}}\bra{\mathbf{0}}}\) can be implemented using a single \((n-1)\)-controlled \(P(N_{v_0}\Delta t\Delta p/(2\pi))\) gate.
Similarly, for the double-separable potential case in \cref{eq:double_separable}, let \(U_{v_1}\) and \(U_{v_2}\) be state-preparation unitaries satisfying \(\ket{v_1}/\sqrt{N_{v_1}\Delta p/(2\pi)} = U_{v_1}\ket{\mathbf{0}}\) and \(\ket{v_2}/\sqrt{N_{v_2}\Delta p/(2\pi)} = U_{v_2}\ket{\mathbf{0}}\), respectively.
Here, \(N_{v_1} = \sum_{j=0}^{2^n-1} |v_1(p_j)|^2\) and \(N_{v_2} = \sum_{j=0}^{2^n-1} |v_2(p_j)|^2\) are the normalization factors of \(\ket{v_1}\) and \(\ket{v_2}\).
Using the first-order Trotterization formula, the operator \(e^{-iV_{\mathrm{nonl}}\Delta t} = e^{i\Delta t (\ket{v_1}\bra{v_1} + \ket{v_2}\bra{v_2})}\) can be implemented as
\begin{multline}
	e^{-iV_{\mathrm{nonl}}\Delta t} \approx U_{v_2} e^{i\Delta t N_{v_2}\Delta p/(2\pi) \ket{\mathbf{0}}\bra{\mathbf{0}}} U_{v_2}^\dagger \times \\
	U_{v_1} e^{i\Delta t N_{v_1}\Delta p/(2\pi) \ket{\mathbf{0}}\bra{\mathbf{0}}} U_{v_1}^\dagger.
\end{multline}

\subsubsection{\label{sec:impl_t_kin}Implementation of \(e^{-i\tilde{T}_{\mathrm{kin}}\Delta t}\)}
Substituting \(p_j = -p_{\mathrm{max}} + j\Delta p\) into the kinetic energy operator \(\tilde{T}_{\mathrm{kin}} = \sum_{j=0}^{2^n-1} (p_j^2/2) \ket{j}\bra{j}\) yields
\begin{equation}
	\tilde{T}_{\mathrm{kin}} = \sum_{j=0}^{2^n-1} \left(\frac{p_{\mathrm{max}}^2}{2} - p_{\mathrm{max}} \Delta p j + \frac{(\Delta p)^2}{2} j^2 \right)\ket{j}\bra{j}.
\end{equation}
Following the procedure in \cref{sec:impl_v_loc}, we expand the quadratic term as
\begin{equation}
	e^{i\tau_2 (\Delta p)^2 j^2} = \prod_{k_0=0}^{n-1} \prod_{k_{1}=0}^{n-1} e^{i2^{k_0+k_{1}} \tau_2 (\Delta p)^2 \delta_{j_{k_0}, 1} \delta_{j_{k_1}, 1}},
\end{equation}
where \(\tau_2 = -\Delta t/2\).
The factor \(e^{i2^{k_0+k_{1}} \tau_2 (\Delta p)^2 \delta_{j_{k_0}, 1} \delta_{j_{k_1}, 1}}\) applies a \(2^{k_0+k_{1}} \tau_2 (\Delta p)^2\) phase shift to the state \(\ket{j}\) if \(j_{k_0} = j_{k_{1}} = 1\), which can be implemented using a controlled-\(P(2^{k_0+k_{1}} \tau_2 (\Delta p)^2)\) gate.
Thus, the full time-evolution operator for the kinetic energy,
\begin{equation}
	e^{-i\tilde{T}_{\mathrm{kin}}\Delta t} = e^{-i\frac{p_{\mathrm{max}}^2}{2}\Delta t} \sum_{j=0}^{2^n-1} e^{i p_{\mathrm{max}} \Delta p j \Delta t} e^{i\tau_2 (\Delta p)^2 j^2}\ket{j}\bra{j},
\end{equation}
is equivalent to a tensor product of single-qubit phase gates, given by
\begin{equation}
	e^{i p_{\mathrm{max}} \Delta p j \Delta t} = \bigotimes_{k=0}^{n-1} P(2^k p_{\mathrm{max}} \Delta p \Delta t),
\end{equation}
followed by a series of controlled-\(P(2^{k_0+k_{1}} \tau_2 (\Delta p)^2)\) gates for \(k_0, k_1 = 0, 1, \ldots, n-1\), up to a global phase.
Note that when \(k_0 = k_1\), the controlled-\(P(2^{k_0+k_{1}} \tau_2 (\Delta p)^2)\) gate reduces to a single-qubit phase gate \(P(2^{k_0+k_{1}} \tau_2 (\Delta p)^2)\).

\subsubsection{\label{sec:impl_time_evolve}Implementation of the Trotter step operator}
\begin{figure*}[t]
	\centering
	\scalebox{0.53}{
		\begin{quantikz}[column sep=0.35cm, row sep=0.2cm]
			\gate{P\left(\phi^{(V_{\mathrm{loc}})}_0\right)}\gategroup[4,steps=1,style={dashed,rounded corners,fill=red!20,inner xsep=1pt},background]{} & \qw\gategroup[4,steps=11,style={dashed,rounded corners,fill=green!20,inner xsep=1pt},background]{} & \qw                              & \qw                              & \gate[2]{U^{[0]\dagger}} & \gate{X} & \ctrl{1}                      & \gate{X} & \gate[2]{U^{[0]}} & \qw                       & \qw                       & \qw                    & \gate{P\left(\phi^{(T_{\mathrm{kin}})}_0\right)}\gategroup[4,steps=7,style={dashed,rounded corners,fill=blue!20,inner xsep=1pt},background]{} & \ctrl{1}                                            & \ctrl{2}                                            & \qw                                                 & \ctrl{3}                                            &                                                     & \qw                                                 & \qw \\
			\gate{P\left(\phi^{(V_{\mathrm{loc}})}_1\right)}                                                                                             & \qw                                                                                                & \qw                              & \gate[3]{U^{[1]\dagger}} & \qw                              & \gate{X} & \ctrl{1}                      & \gate{X} & \qw                       & \gate[3]{U^{[1]}} & \qw                       & \qw                    & \gate{P\left(\phi^{(T_{\mathrm{kin}})}_1\right)}                                                                                              & \gate{P\left(\phi^{(T_{\mathrm{kin}})}_{01}\right)} & \qw                                                 & \ctrl{1}                                            & \qw                                                 & \ctrl{2}                                            & \qw                                                 & \qw \\
			\gate{P\left(\phi^{(V_{\mathrm{loc}})}_2\right)}                                                                                             & \qw                                                                                                & \gate[2]{U^{[2]\dagger}} & \qw                              & \qw                              & \gate{X} & \ctrl{1}                      & \gate{X} & \qw                       & \qw                       & \gate[2]{U^{[2]}} & \qw                    & \gate{P\left(\phi^{(T_{\mathrm{kin}})}_2\right)}                                                                                              & \qw                                                 & \gate{P\left(\phi^{(T_{\mathrm{kin}})}_{02}\right)} & \gate{P\left(\phi^{(T_{\mathrm{kin}})}_{12}\right)} & \qw                                                 & \qw                                                 & \ctrl{1}                                            & \qw \\
			\gate{P\left(\phi^{(V_{\mathrm{loc}})}_3\right)}                                                                                             & \gate{U^{[3]\dagger}}                                                                      & \qw                              & \qw                              & \qw                              & \gate{X} & \gate{P(N_{v_0} \Delta t\Delta p /(2\pi))} & \gate{X} & \qw                       & \qw                       & \qw                       & \gate{U^{[3]}} & \gate{P\left(\phi^{(T_{\mathrm{kin}})}_3\right)}                                                                                              & \qw                                                 & \qw                                                 & \qw                                                 & \gate{P\left(\phi^{(T_{\mathrm{kin}})}_{03}\right)} & \gate{P\left(\phi^{(T_{\mathrm{kin}})}_{13}\right)} & \gate{P\left(\phi^{(T_{\mathrm{kin}})}_{23}\right)} & \qw
		\end{quantikz}
	}
	\caption{An example 4-qubit quantum circuit diagram implementing a single Trotter step operator for a one-dimensional hydrogen atom interacting with an ultrashort laser pulse using a full-rank MPS representation of the single non-local separable potential. The first segment (filled with red) of the circuit implements \(e^{-iV_{\mathrm{loc}}(t)\Delta t}\) as a tensor product of single-qubit phase gates, the second segment (filled with green) implements \(e^{-iV_{\mathrm{nonl}}\Delta t}\) using a state-preparation unitary \(U_{v_0}\) and an \((n-1)\)-controlled phase gate, and the third segment (filled with blue) implements \(e^{-i\tilde{T}_{\mathrm{kin}}\Delta t}\) as a tensor product of single-qubit phase gates followed by a series of controlled-phase gates. \(\phi^{(V_{\mathrm{loc}})}_k=-2^k \tau_1 \Delta p\) and \(\phi^{(T_{\mathrm{kin}})}_k=2^k p_{\mathrm{max}} \Delta p \Delta t + 2^{2k} \tau_2 (\Delta p)^2\) are the phase angles for the \(k\)-th qubit, and \(\phi^{(T_{\mathrm{kin}})}_{kl}=2^{k+l+1} \tau_2 (\Delta p)^2\) is the phase angle for the controlled-phase gate between the \(k\)-th and \(l\)-th qubits. The state-preparation unitaries \(U^{[k]}\) are implemented using the MPS representation of \(\ket{v_0}\).
	}\label{fig:qiskit_circuit}
\end{figure*}
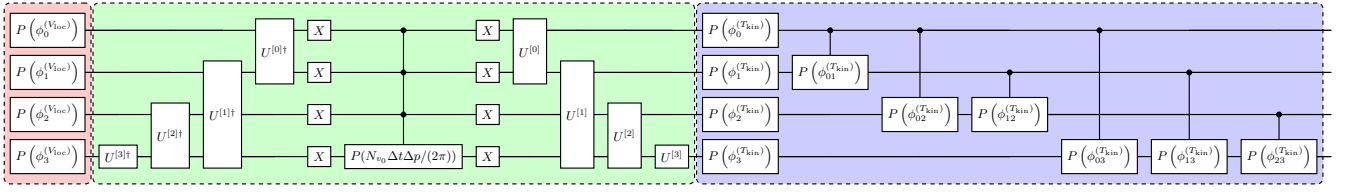

By combining the aforementioned implementations of \(e^{-iV_{\mathrm{loc}}(t)\Delta t}\), \(e^{-iV_{\mathrm{nonl}}\Delta t}\), and \(e^{-i\tilde{T}_{\mathrm{kin}}\Delta t}\), we can implement the Trotter step operator given by \cref{eq:split_operator_1dh} as the quantum circuit shown in \cref{fig:qiskit_circuit}.

\subsection{\label{sec:potential_bonddim}Bond dimension of MPS representation of separable potentials}
\begin{figure}
	\centering
	\includegraphics[width=\linewidth]{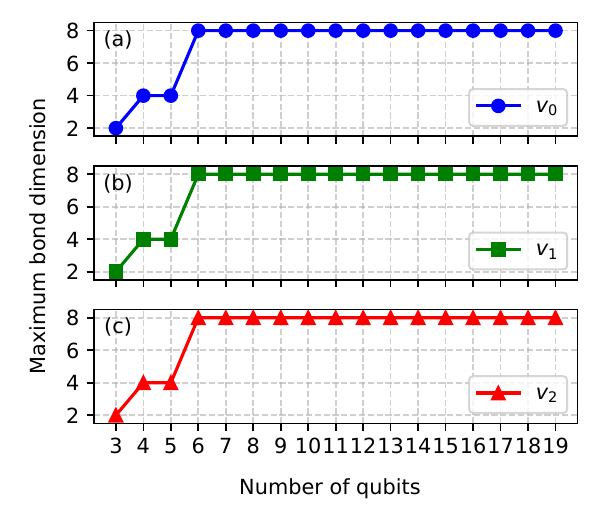}
	\caption{Maximum bond dimension of the MPS representations of (a) \(\overline{\ket{v_0}}\), (b) \(\overline{\ket{v_1}}\), and (c) \(\overline{\ket{v_2}}\) for each number of qubits.}\label{fig:bonddim}
\end{figure}

We first evaluated the maximum bond dimension required to approximate each separable-potential factor to a fixed truncation accuracy as a function of the number of grid qubits.
We converted a normalized state vector \(\overline{\ket{v_k}} = \ket{v_k}/\sqrt{N_{v_k}}\) into an \(n\)-site MPS with a singular value cutoff of \(10^{-15}\) using the Quimb library~\cite{Gray2018-ac}, recording the maximum bond dimension of the resulting MPS for each number of qubits \(n\) ranging from 3 to 19.
As shown in \cref{fig:bonddim}(a), the maximum bond dimension for the potential $v_0$ saturated at 8, even when the number of qubits was increased beyond 6.
A similar trend was observed for \(v_1\) and \(v_2\) as shown in Figs.~\ref{fig:bonddim}(b) and \ref{fig:bonddim}(c), respectively.
\begin{figure}
	\centering
	\includegraphics[width=\linewidth]{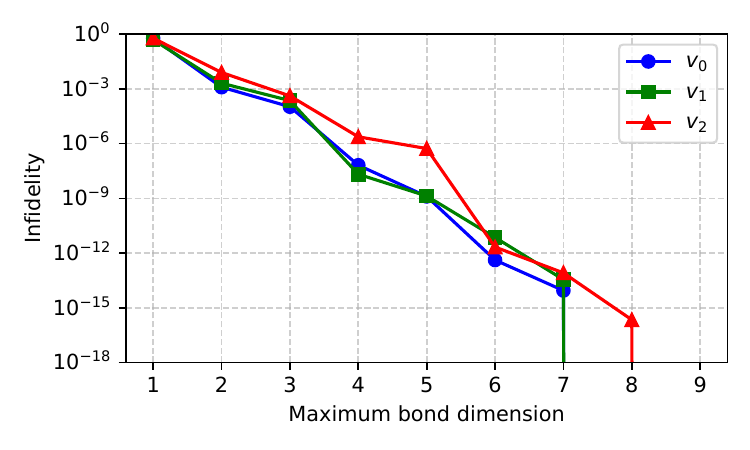}
	\caption{Infidelity of the MPS representations of \(\overline{\ket{v_0}}\), \(\overline{\ket{v_1}}\), and \(\overline{\ket{v_2}}\) for each maximum bond dimension from 1 to 9. The infidelity drops to zero at a maximum bond dimension of 8 or more for \(\overline{\ket{v_0}}\) and \(\overline{\ket{v_1}}\), and of 9 for \(\overline{\ket{v_2}}\).}\label{fig:infidelity_vs_bonddim}
\end{figure}
\Cref{fig:infidelity_vs_bonddim} illustrates the infidelity of the MPS representations for \(\overline{\ket{v_0}}\), \(\overline{\ket{v_1}}\), and \(\overline{\ket{v_2}}\) for each maximum bond dimension. The infidelity decreases exponentially as the maximum bond dimension increases, dropping below \(10^{-15}\) for bond dimensions of 8 or greater.
This saturation of the maximum bond dimension occurs because the separable potentials \(v_0\), \(v_1\), and \(v_2\) are smooth functions in momentum space, allowing them to be well approximated by low-rank MPS representations.
Specifically, the formula for \(v_0\) in \cref{eq:v0} consists of a product of a polynomial and a Gaussian function. While \(v_1\) and \(v_2\) are initially expressed as linear combinations of such products, the specific choice of coefficients \(\alpha_{11}, \alpha_{12}, \alpha_{21}\), and \(\alpha_{22}\) in \cref{eq:a_variables_def} reduces them to a simple product of a polynomial and a Gaussian function as well.
It has been proven that a polynomial of degree \(d\) evaluated on an equidistant grid can be exactly represented by an MPS with a maximum bond dimension of \(d+1\)~\cite{Grasedyck2010-nw}.
Furthermore, when the number of qubits is increased to represent a smooth, differentiable real-valued function such as a Gaussian, the additional entanglement entropy decreases exponentially~\cite{Garcia-Ripoll2021-wa,Iaconis2024-rd}.
The upper bound on the maximum bond dimension for an MPS representing the product of two functions is given by the product of their individual maximum bond dimensions.
Since the degrees of the polynomials in \(v_0\), \(v_1\), and \(v_2\) are at most 3, the maximum bond dimension of the resulting MPS is primarily dominated by the Gaussian function (see \cref{app:schmidt} for details).

\subsection{\label{sec:dependence_bonddim}Dependence of the dynamics on the MPS bond dimension}
\begin{figure}
	\centering
	\includegraphics[width=\linewidth]{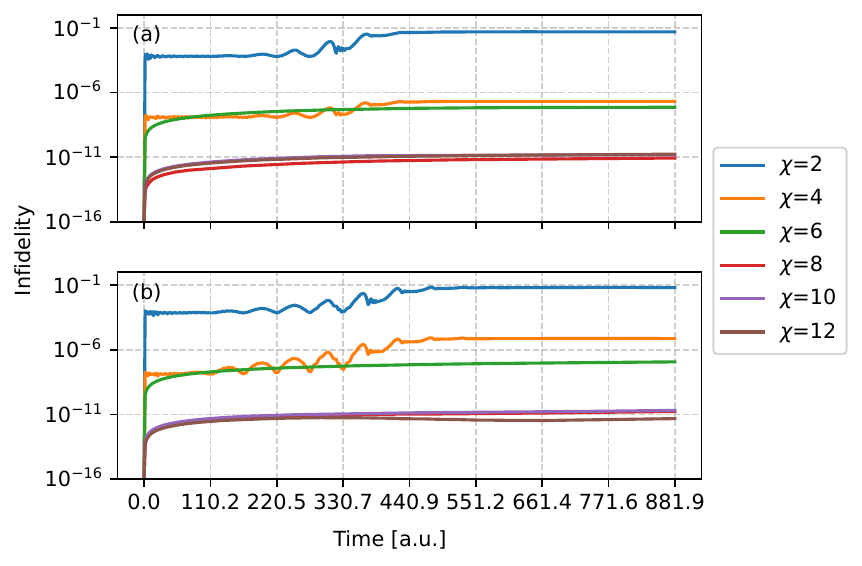}
	\caption{The temporal evolution of infidelity when using the MPS-encoded non-local potentials (a) \(V(p,p^\prime) = -v_0(p)v_0^*(p^\prime)\) and (b) \(V(p,p^\prime) = -v_1(p)v_1^*(p^\prime) -v_2(p)v_2^*(p^\prime)\), compared with the split-operator method over the same 10,000 time steps for each bond dimension.}\label{fig:changechi}
\end{figure}

Next, we constructed a 10-qubit quantum circuit to perform the time evolution of the ground-state wave function over 10,000 time steps.
This was achieved using the Trotter step operator given by \cref{eq:split_operator_1dh}, with the non-local potential \(V(p,p^\prime) = -v_0(p)v_0^*(p^\prime)\) represented by an MPS. We evaluated the circuit performance across different maximum bond dimensions (\(\chi = 2,4,6,8,10,\) and \(12\)).
The simulation parameters were identical to those used in \cref{fig:survival}.
As shown in \cref{fig:changechi}(a), we calculated the infidelity against the reference state obtained via the standard split-operator method over the same 10,000 time steps.
The infidelity is defined as \(1 - |\braket{\psi_{\mathrm{SO}}(t)}{\psi_{\mathrm{MPS}}(t)}|^2\), where \(|\psi_{\mathrm{SO}}(t)\rangle\) is the wave function obtained by the split-operator method and \(|\psi_{\mathrm{MPS}}(t)\rangle\) is the wave function produced by the quantum circuit using the MPS-encoded non-local potential.
Comparing the final states after the time evolution, we observed that the infidelity decreased monotonically as \(\chi\) was increased from 2 to 8.
Increasing the bond dimension beyond 8 yielded no further improvement. This behavior is expected because, as discussed in \cref{sec:potential_bonddim}, the exact MPS representation of the potential for 10 qubits saturates at a maximum bond dimension of 8.
We performed analogous calculations for the double-separable non-local potential using MPS-encoded \(v_1\) and \(v_2\), yielding similar results [\cref{fig:changechi}(b)].

\subsection{\label{sec:depth}Circuit depth}
Here, we investigated the circuit depth of the single Trotter step operator for the one-dimensional hydrogen atom interacting with an ultrashort laser pulse, utilizing the MPS representation of non-local separable potentials.
In \cref{sec:generalimpl}, we established the circuit depth for a general system with \(n_{\mathrm{atom}}\) atoms, a maximum of \(L\) projector functions across all atoms, \(n\) qubits, and a maximum MPS bond dimension \(\chi\).
For the one-dimensional hydrogen atom case, we have \(n_{\mathrm{atom}} = 1\), with \(L = 1\) for the single-separable potential and \(L = 2\) for the double-separable potential.
Because the simulation is performed in the momentum representation, the quantum Fourier transform is not required.
Consequently, the exact gate count for a single Trotter step is bounded by approximately \(O(8n\chi^2+2n^2)\) for the single-separable case and \(O(16n\chi^2+2n^2)\) for the double-separable case.

\begin{figure}
	\centering
	\includegraphics[width=\linewidth]{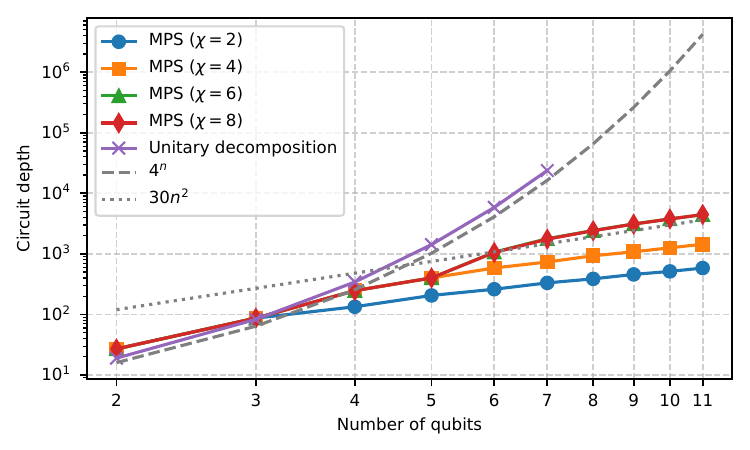}
	\caption{Circuit depth of a single Trotter step for the one-dimensional hydrogen atom interacting with an ultrashort laser pulse, utilizing the MPS-encoded single-separable non-local potential with maximum bond dimensions \(\chi = 2,4,6,\) and \(8\) for different numbers of qubits \(n\). For reference, the depth of the exact unitary decomposition without the MPS approximation is also shown (labeled as ``Unitary decomposition''). Data for the exact unitary decomposition are only plotted up to \(n \leq 7\) due to the exponential runtime of the synthesis algorithm.
	The dashed line represents the plot of \(4^n\) and the dotted line represents the plot of \(30n^2\).
	}\label{fig:circuit_depth}
\end{figure}

To empirically evaluate this, we constructed the corresponding quantum circuit for the single-separable potential case and measured its depth using Qiskit (version 1.0.2)~\cite{Javadi-Abhari2024-xr}.
The evaluated circuit corresponds to the first Trotter step applied to the initial ground-state wave function at \(t = 0\), and the parameters of the laser pulse are identical to those used in \cref{fig:survival}.
We optimized the circuit using the Qiskit transpiler with optimization level 2, and decomposed the circuit into a basis gate set \(\{R_X, R_Z, \text{CNOT}\}\), where \(R_X\) and \(R_Z\) are single-qubit rotation gates around the X and Z axes, respectively, and CNOT is a controlled-NOT gate. We recorded the circuit depth for varying numbers of qubits \(n\), ranging from 2 to 11.
The results are shown in \cref{fig:circuit_depth}.
For comparison, the figure also includes the circuit depth of the Trotter step implemented without the MPS approximation.
In this reference case, the non-local potential propagator \(U_{\mathrm{nonl}}\) is treated as a dense unitary operator and decomposed using Qiskit's unitary synthesis algorithm.
As illustrated, the circuit depth for the exact unitary decomposition scales exponentially with the number of qubits as \(O(4^n)\), whereas the depth of the MPS-based circuit scales polynomially as \(O(n^2)\) for a fixed bond dimension \(\chi\).
For instance, at \(n=7\), the depth of the exact unitary decomposition circuit was 23,775, whereas the depth of the proposed circuit using the MPS with a maximum bond dimension of \(\chi=8\) was only 1,747. This demonstrates that the conventional unitary decomposition results in a circuit more than 13 times deeper than our proposed MPS representation of non-local potentials.
The empirical results are consistent with the theoretical analysis of the circuit depth of \(O(8n\chi^2+2n^2)\) for the single-separable potential case, confirming that the proposed method is effective in reducing the circuit depth.

\section{Conclusion}
In this study, we proposed a method for representing non-local potentials using MPS in first-quantized quantum simulations.
Our approach first represents the wave function using binary encoding, and then splits the time-evolution operator into a kinetic part \(e^{-iT_{\mathrm{kin}}\Delta t}\) and a potential part \(e^{-iV\Delta t}\) via first-order Trotterization.
The potential can be divided into local and non-local parts as \(V = V_{\mathrm{loc}} + V_{\mathrm{nonl}}\), where the non-local component is efficiently represented using an MPS.
Since an MPS can be directly mapped to a quantum circuit, this allows for the seamless implementation of the non-local potential.
Let \(n_{\mathrm{atom}}\) be the number of atoms in the system, \(L\) be the maximum number of projector functions across all atoms, \(n\) be the number of qubits, and \(\chi\) be the maximum bond dimension of all MPSs representing the non-local potentials.
Our method achieves a circuit depth of \(O(\deg(V_{\mathrm{loc}}) n^{\deg(V_{\mathrm{loc}})} + 8Lnn_{\mathrm{atom}}\chi^2 + 2n^2)\) for the quantum circuit implementing the single Trotter step operator, assuming \(V_{\mathrm{loc}}\) is polynomial.

We tested this method by simulating the ionization dynamics of a one-dimensional hydrogen atom subject to an ultrashort laser pulse, utilizing a non-local separable potential.
The maximum bond dimension required for the MPS representation saturated at 8, remaining independent of the number of grid qubits.
We also confirmed the effectiveness of the proposed method in reducing the circuit depth. Specifically, for \(n = 7\) qubits, the depth of the quantum circuit implementing a single Trotter step with an MPS-encoded single non-local separable potential with a bond dimension of 8 was reduced to under 8\% of the depth required for a standard direct unitary decomposition of the non-local potential propagator.

Although we have only demonstrated the effectiveness of the proposed method using a one-dimensional single-electron system, the proposed method is applicable to three-dimensional multi-electron systems. A previous study has shown that representing Gaussian-type molecular orbitals with MPS achieves high accuracy at a low bond dimension~\cite{Huggins2025-bv}. This suggests that our proposed method will remain effective in reducing the circuit depth when simulating three-dimensional multi-electron dynamics employing non-local pseudopotentials.

\begin{acknowledgments}
	This research was supported by the Forefront Physics and Mathematics Program to Drive Transformation (FoPM), a World-leading Innovative Graduate Study (WINGS) Program, the University of Tokyo.
	This work was supported by JSPS KAKENHI Grant Number 25KJ0799 and the Center of Innovation for Sustainable Quantum AI (JST Grant Number JPMJPF2221).
	This study was carried out using the TSUBAME4.0 supercomputer at the Institute of Science Tokyo.
\end{acknowledgments}

\appendix

\section{\label{app:derivation}Derivation of the non-local separable potential for the ground state}
To remove the singularity of the Coulomb potential in the one-dimensional hydrogen atom model, we introduce the soft Coulomb potential,
\begin{equation}
	V_{\mathrm{soft}}(x) = - \frac{1}{\sqrt{x^2 + \epsilon^2}}.
\end{equation}
The time-independent Schr\"odinger equation for this model is given by
\begin{equation}\label{eq:1d_se_soft}
	- \frac{1}{2} \frac{d^2\psi}{dx^2} - \frac{1}{\sqrt{x^2 + \epsilon^2}}\psi = E\psi.
\end{equation}
We aim to replace this soft Coulomb potential with a separable non-local potential:
\begin{equation}\label{eq:non-local_pot_def_1}
	V(p, p^\prime) = - v(p) v^*(p^\prime).
\end{equation}
We require that the corresponding Schr\"odinger equation in the momentum representation preserve the exact ground-state solution \(\phi_0\) of \cref{eq:1d_se_soft}:
\begin{equation}\label{eq:1d_se_non-local}
	\frac{p^2}{2}\phi(p) + \int \frac{dp^\prime}{2 \pi} V(p, p^\prime)\phi(p^{\prime}) = E\phi(p).
\end{equation}
By defining
\begin{equation}\label{eq:a_1d}
	a = \int \frac{d p^\prime}{2 \pi} v^*(p^\prime) \phi_0(p^\prime),
\end{equation}
we obtain
\begin{equation}
	\left(E_0 - \frac{p^2}{2}\right) \phi_0(p) = -a v(p).
\end{equation}
This leads to
\begin{equation}\label{eq:separable_pot_v_def}
	v(p) = \frac{p^2 - 2E_0}{2a} \phi_0(p).
\end{equation}
Substituting this expression into \cref{eq:a_1d}, we obtain
\begin{equation}
	a = \int \frac{d p^\prime}{2 \pi} v^*(p^\prime) \phi_0(p^\prime) = \int \frac{d p^\prime}{2 \pi} \frac{p^{\prime 2} - 2E_0}{2a} |\phi_0(p^\prime)|^2.
\end{equation}
Assuming \(a>0\), this can be rewritten as
\begin{equation}\label{eq:def_a}
	a = \sqrt{\left(\int \frac{d p^\prime}{2 \pi} \frac{p^{\prime 2}}{2}|\phi_0(p^\prime)|^2\right) - E_0}.
\end{equation}
Evaluating this expression determines the functional form of \(v(p)\), which in turn yields the non-local potential \(V(p, p^\prime)\).

The analytical solution for the soft Coulomb potential involves an infinite series~\cite{Li2021-mq}, making the integration difficult. Therefore, we adopted an approximate solution~\cite{Grasselli2017-xx}.
As an ansatz for the ground state, we introduce
\begin{equation}\label{eq:ansatz_gs_x}
	\phi_0 (x) = \left(\frac{\beta_0}{\pi}\right)^{1/4} e^{-\beta_0 x^2/2},
\end{equation}
and numerically determine the parameter \(\beta_0\) that minimizes the corresponding ground-state energy:
\begin{equation}
	E_0 = \frac{\beta_0}{4} - \sqrt{\frac{\beta_0}{\pi}} e^{\epsilon^2\beta_0/2} K_0\left(\frac{\epsilon^2 \beta_0}{2}\right),
\end{equation}
where \(K_0(x)\) is the zeroth-order modified Bessel function of the second kind.
We then construct the non-local potential defined in \cref{eq:non-local_pot_def_1} such that it yields this optimized approximate wave function as its exact solution.
First, we perform the Fourier transform of \cref{eq:ansatz_gs_x}:
\begin{equation}
	\phi_0 (p) = \int_{-\infty}^\infty dx \phi_0 (x) e^{-ipx} = \left(\frac{4\pi}{\beta_0}\right)^{1/4} e^{-\frac{p^2}{2\beta_0}}.
\end{equation}
Substituting this into \cref{eq:def_a} and noting that
\begin{equation}
	\int_{-\infty}^{\infty} p^2 e^{-p^2/\beta_0} dp = \frac{\beta_0}{2} \sqrt{\pi\beta_0},
\end{equation}
we obtain
\begin{equation}
	a = \sqrt{\frac{\beta_0}{4} - E_0}.
\end{equation}
Substituting this back into \cref{eq:separable_pot_v_def} yields
\begin{equation}
	v(p) = \frac{p^2 -2E_0}{ \sqrt{\beta_0 - 4E_0}} \phi_0(p).
\end{equation}

\section{\label{app:derivation2}Derivation of the non-local separable potential for the ground state and the first excited state}
Here, we aim to replace the soft Coulomb potential with a non-local potential of the form:
\begin{equation}\label{eq:non-local_pot_def_2}
	V(p, p^\prime) = - v_1(p) v_1^*(p^\prime) - v_2(p) v_2^*(p^\prime).
\end{equation}
We require that the corresponding Schr\"odinger equation in the momentum representation possess exactly the same solutions for the ground state \(\phi_0\) and the first excited state \(\phi_1\) as those of \cref{eq:1d_se_soft}:
\begin{equation}\label{eq:1d_se_non-local_2}
	\frac{p^2}{2}\phi(p) + \int \frac{dp^\prime}{2 \pi} V(p, p^\prime)\phi(p^{\prime}) = E\phi(p).
\end{equation}
By defining
\begin{equation}
	a_{jn} = \int \frac{d p^\prime}{2 \pi} v_n^*(p^\prime) \phi_{j-1}(p^\prime) \quad (j, n \in \{1, 2\}),
\end{equation}
we obtain the following relations from \cref{eq:1d_se_non-local_2}:
\begin{align}
	\left(E_0 - \frac{p^2}{2}\right) \phi_0(p) + a_{11}v_1(p) + a_{12}v_2(p) & = 0, \\
	\left(E_1 - \frac{p^2}{2}\right) \phi_1(p) + a_{21}v_1(p) + a_{22}v_2(p) & = 0.
\end{align}
This system can be expressed in matrix form as
\begin{equation}
	\begin{pmatrix}
		\left(E_0 - \frac{p^2}{2}\right) \phi_0(p) \\ \left(E_1 - \frac{p^2}{2}\right) \phi_1(p)
	\end{pmatrix}
				=-\begin{pmatrix}
						a_{11} & a_{12} \\ a_{21} & a_{22}
					\end{pmatrix}\begin{pmatrix}
									v_1(p) \\ v_2(p)
								\end{pmatrix},
\end{equation}
or equivalently,
\begin{equation}
	\bm{\Phi} = -A\bm{v},
\end{equation}
where \(\bm{\Phi}\) and \(\bm{v}\) denote the column vectors on the left- and right-hand sides of the preceding matrix equation, respectively. Thus, if the matrix \(A\) is invertible, we can express \(\bm{v}\) as
\begin{equation}
	\bm{v} = -A^{-1}\bm{\Phi}.
\end{equation}
This allows us to determine the components of \(\bm{v}\) by working backward from the target wave functions that must satisfy the Schr\"odinger equation.
We then consider the matrix
\begin{equation}
	\Gamma = AA^T.
\end{equation}
Its elements are given by
\begin{widetext}
	\begin{equation}
		\gamma_{ij} = -\int_{-\infty}^\infty \frac{dp}{2\pi} \phi_{i-1}^*(p) \left(E_{j-1} - \frac{p^2}{2}\right)\phi_{j-1}(p) = -\int_{-\infty}^\infty dx \phi_{i-1}^*(x) V_{\mathrm{soft}}(x)\phi_{j-1}(x),
	\end{equation}
\end{widetext}
which can be computed directly. Since
\begin{align}
	AA^T & = \begin{pmatrix}
		         a_{11} & a_{12} \\
		         a_{21} & a_{22}
	         \end{pmatrix}\begin{pmatrix}
		                      a_{11} & a_{21} \\
		                      a_{12} & a_{22}
	                      \end{pmatrix} \nonumber                                                                                           \\
	     & =\begin{pmatrix}
		        a_{11}^2+a_{12}^2                               & a_{11}a_{21}+a_{22}a_{12} \\
		        a_{11}a_{21}+a_{22}a_{12} & a_{21}^2+a_{22}^2
	        \end{pmatrix} \nonumber \\
	     & =\begin{pmatrix}
		        \gamma_{11} & \gamma_{12} \\
		        \gamma_{21} & \gamma_{22}
	        \end{pmatrix},
\end{align}
we obtain the following system of equations:
\begin{equation}
	\begin{dcases}
		a_{11}^2 + a_{12}^2 = \gamma_{11} \\
		a_{21}^2 + a_{22}^2 = \gamma_{22} \\
		a_{11}a_{21}+a_{22}a_{12} = \gamma_{12}
	\end{dcases}.\label{eq:A_mat_eq}
\end{equation}
We parametrize the elements of \(A\) as
\begin{equation}
	\begin{aligned}
	&a_{11} = \sqrt{\gamma_{11}} \cos \theta, \quad a_{12} = \sqrt{\gamma_{11}} \sin \theta,\\
	&a_{21} = \sqrt{\gamma_{22}} \cos \varphi, \quad a_{22} = \sqrt{\gamma_{22}} \sin \varphi.
	\end{aligned}
\end{equation}
From the third equation of \cref{eq:A_mat_eq}, we obtain
\begin{equation}
	\sqrt{\gamma_{11}\gamma_{22}} \cos (\varphi - \theta) = \gamma_{12}.
\end{equation}
Because \(|\gamma_{12}| \leq \sqrt{\gamma_{11}\gamma_{22}}\) holds by the Cauchy--Schwarz inequality, we define \(\kappa\) such that
\begin{equation}
	\sin \kappa = \frac{\gamma_{12}}{\sqrt{\gamma_{11}\gamma_{22}}},
\end{equation}
which gives \(\cos(\varphi - \theta) = \sin \kappa\). This is satisfied by \(\varphi = \theta + \pi/2 - \kappa\), and therefore
\begin{equation}
	A = \begin{pmatrix}
		\sqrt{\gamma_{11}} \cos \theta & \sqrt{\gamma_{11}} \sin \theta \\
		\sqrt{\gamma_{22}} \sin (\kappa - \theta) & \sqrt{\gamma_{22}} \cos (\kappa - \theta)
	\end{pmatrix}.
\end{equation}
Assuming \(\det A = \sqrt{\gamma_{11}\gamma_{22}} \cos \kappa \neq 0\), we define \(\alpha_{ij}\) as the elements of \(-A^{-1}\):
\begin{equation}
	-A^{-1} =
	-\frac{1}{\det A}\begin{pmatrix}
		a_{22} & -a_{12} \\
		-a_{21} & a_{11}
	\end{pmatrix}
	=\begin{pmatrix}
		\alpha_{11} & \alpha_{12} \\
		\alpha_{21} & \alpha_{22}
	\end{pmatrix},
\end{equation}
which gives the general solution~\cite{Tetchou-Nganso2011-ie}
\begin{equation}
	\begin{aligned}
		 & \alpha_{11}=-\frac{\cos (\kappa-\theta)}{\sqrt{\gamma_{11}} \cos \kappa}, \quad \alpha_{12}=\frac{\sin \theta}{\sqrt{\gamma_{22}} \cos \kappa}, \\
		 & \alpha_{21}=\frac{\sin (\kappa-\theta)}{\sqrt{\gamma_{11}} \cos \kappa}, \quad \alpha_{22}=-\frac{\cos \theta}{\sqrt{\gamma_{22}} \cos \kappa},
	\end{aligned}
\end{equation}
where \(\kappa\) has two possible values: \(\kappa_1=\arcsin \left(\gamma_{12} / \sqrt{\gamma_{11} \gamma_{22}}\right)\) and \(\kappa_2=\pi- \arcsin \left(\gamma_{12} / \sqrt{\gamma_{11} \gamma_{22}}\right)\).
In this study, we chose \(\kappa = \kappa_2\) and \(\theta = 0\), which yields
\begin{equation}
	\begin{aligned}
		 & \alpha_{11}=-\frac{1}{\sqrt{\gamma_{11}}}, \quad \alpha_{12}=0,                                                             \\
		 & \alpha_{21}=\frac{\tan{\kappa_2}}{\sqrt{\gamma_{11}}}, \quad \alpha_{22}=-\frac{1}{\sqrt{\gamma_{22}} \cos \kappa_2}.
	\end{aligned}
\end{equation}
Next, we determine the approximate solution \(\phi_1\) for the first excited state of the Schr\"odinger equation.
As an ansatz, we propose
\begin{equation}\label{eq:ansatz_es_x}
	\phi_1 (x) = \left(\frac{\beta_1}{\pi}\right)^{1/4} \sqrt{2\beta_1} x e^{-\beta_1 x^2/2},
\end{equation}
and numerically find the parameter \(\beta_1\) that minimizes the energy:
\begin{equation}
	E_1 = \frac{3\beta_1}{4} - \sqrt{\beta_1} U\left(\frac{1}{2},0,\epsilon^2\beta_1\right),
\end{equation}
where \(U(a,b,x)\) is Tricomi's confluent hypergeometric function.
Taking the Fourier transform of \cref{eq:ansatz_es_x}, we obtain
\begin{align}
	\phi_1 (p) = -2ip \left( \frac{\pi}{\beta_1^3} \right)^{1/4} e^{-\frac{p^2}{2\beta_1}}.
\end{align}
We then evaluate each element of \(\Gamma\) as follows:
\begin{align}
	\gamma_{11} & = -\int_{-\infty}^\infty dx \phi_0^*(x) V_{\mathrm{soft}}(x)\phi_0(x) \nonumber                                      \\
	            & = \int_{-\infty}^\infty dx \left(\frac{\beta_0}{\pi}\right)^{1/2}  \frac{e^{-\beta_0 x^2}}{\sqrt{x^2+\epsilon^2}} \nonumber \\
	            & = \left(\frac{\beta_0}{\pi}\right)^{1/2} e^{\epsilon^2\beta_0/2} K_0\left(\frac{\epsilon^2 \beta_0}{2}\right).\label{eq:gamma11}
\end{align}
In the above derivation, we used the following integral:
\begin{equation}
	\mathcal{I}(\beta) = \int_0^\infty \frac{e^{-\beta x^2}}{\sqrt{x^2+\epsilon^2}} dx = \frac{1}{2} e^{\frac{\beta \epsilon^2}{2}} K_0\left(\frac{\beta \epsilon^2}{2}\right).
\end{equation}
Differentiating both sides with respect to \(\beta\), we obtain
\begin{equation}
	\frac{d\mathcal{I}}{d\beta} = -\int_0^\infty \frac{x^2 e^{-\beta x^2}}{\sqrt{x^2+\epsilon^2}} dx = \frac{\epsilon^2}{4} e^z \left[K_0(z) - K_1(z)\right],
\end{equation}
where \(z = \beta \epsilon^2/2\) and \(K_1(x)\) is the first-order modified Bessel function of the second kind.
Using this result, we evaluate \(\gamma_{22}\) as follows:
\begin{align}
	\gamma_{22} & = -\int_{-\infty}^\infty dx \phi_1^*(x) V_{\mathrm{soft}}(x)\phi_1(x) \nonumber                                                                                                                                \\
	            & = \int_{-\infty}^\infty dx \left(\frac{\beta_1}{\pi}\right)^{1/2} 2\beta_1 \frac{x^2 e^{-\beta_1 x^2}}{\sqrt{x^2+\epsilon^2}} \nonumber                                                                               \\
	            & \begin{multlined}
					= 4 \left(\frac{\beta_1^3}{\pi}\right)^{1/2} \frac{\epsilon^2}{4} e^{\frac{\beta_1 \epsilon^2}{2}} \times \\
					\left[K_1\left(\frac{\beta_1 \epsilon^2}{2}\right) - K_0\left(\frac{\beta_1 \epsilon^2}{2}\right)\right].
				\end{multlined}\label{eq:gamma22}
\end{align}
Finally, for \(\gamma_{12}\):
\begin{align}
	\gamma_{12} & = -\int_{-\infty}^\infty dx \phi_0^*(x) V_{\mathrm{soft}}(x)\phi_1(x) \nonumber                                                                                                                 \\
	            & \begin{multlined}
					= \int_{-\infty}^\infty dx \left(\frac{\beta_0}{\pi}\right)^{1/4}\left(\frac{\beta_1}{\pi}\right)^{1/4} \times \\
					\sqrt{2\beta_1} x \frac{e^{-\beta_0 x^2/2}e^{-\beta_1 x^2/2}}{\sqrt{x^2+\epsilon^2}}
				\end{multlined} \nonumber \\
	            & = 0 \quad \text{(odd function)}.\label{eq:gamma12}
\end{align}
Therefore, we set the variables \(\alpha_{11}, \alpha_{12}, \alpha_{21}\), and \(\alpha_{22}\) as follows:
\begin{equation}
	\alpha_{11}=-\frac{1}{\sqrt{\gamma_{11}}}, \quad \alpha_{12}=0, \quad \alpha_{21}=0, \quad \alpha_{22}=\frac{1}{\sqrt{\gamma_{22}}}.
\end{equation}

\section{\label{app:schmidt}Schmidt decomposition of \(\overline{\ket{v_0}}\), \(\overline{\ket{v_1}}\), and \(\overline{\ket{v_2}}\)}
\begin{figure}[h]
	\centering
	\includegraphics[width=\linewidth]{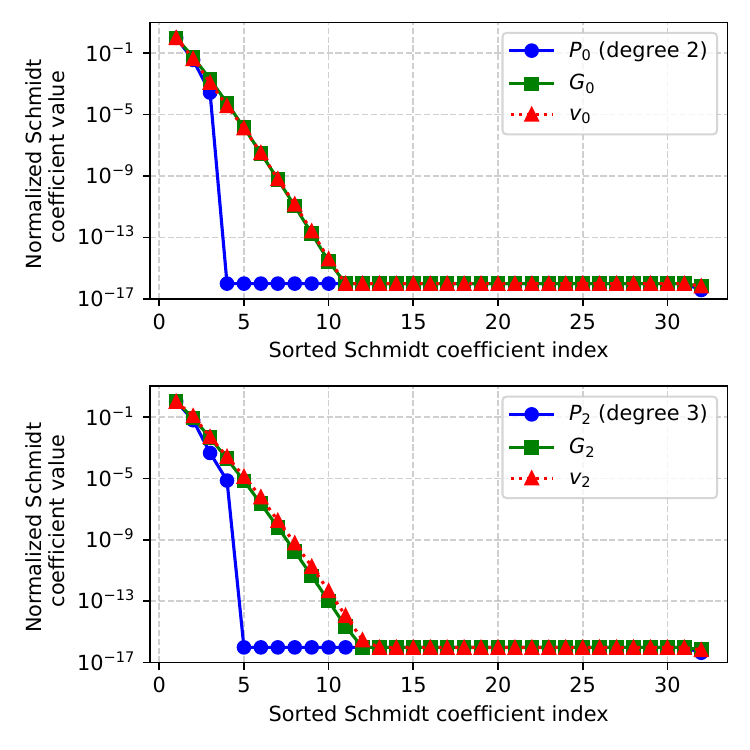}
	\caption{Sorted Schmidt coefficients for the 5-qubit/5-qubit bipartition of \(\ket{P_k}\), \(\ket{G_k}\), and \(\overline{\ket{v_k}}\) for \(k=0,2\). Here, we omit \(k=1\) because \(\ket{P_1}\) and \(\ket{G_1}\) are the same as \(\ket{P_0}\) and \(\ket{G_0}\), respectively.
	}\label{fig:schmidt_vals}
\end{figure}

Here, we perform a Schmidt decomposition on the normalized 10-qubit state vectors \(\overline{\ket{v_0}}\), \(\overline{\ket{v_1}}\), and \(\overline{\ket{v_2}}\) with respect to a 5-qubit/5-qubit bipartition and plot the sorted Schmidt coefficients in \cref{fig:schmidt_vals}.
The Schmidt coefficients \(\lambda_i\) were normalized such that \(\sum_i \lambda_i^2 = 1\). We also factor each \(v_k\) (\(k=0,1,2\)) into a polynomial component \(P_k(p)\) and a Gaussian component \(G_k(p)\), such that \(v_k(p)=P_k(p)G_k(p)\):
\begin{align}
	P_k(p) & \propto
	\begin{cases}
		p^2 - 2E_0          & (k=0,1) \\
		p(p^2 - 2E_1) & (k=2)
	\end{cases}, \\
	G_k(p) & \propto
	\begin{cases}
		e^{-p^2/(2\beta_0)} & (k=0,1) \\
		e^{-p^2/(2\beta_1)} & (k=2)
	\end{cases},
\end{align}
and normalize them to satisfy \(\sum_j |P_k(p_j)|^2 = 1\) and \(\sum_j |G_k(p_j)|^2 = 1\).
The Schmidt coefficients of the quantum states corresponding to the polynomial component, \(\ket{P_k}=\sum_j P_k(p_j) \ket{j}\), and the Gaussian component, \(\ket{G_k}=\sum_j G_k(p_j) \ket{j}\), are also plotted in \cref{fig:schmidt_vals}.
As discussed in \cref{sec:potential_bonddim}, the Schmidt coefficients of the polynomial component remain non-zero (above the machine precision of \(\sim 10^{-16}\)) only for the first \(\deg(P_k) + 1\) indices.
In contrast, those of the Gaussian component decay exponentially but maintain non-zero values over a broader range.
This indicates that the required bond dimension for the MPS representation of \(\overline{\ket{v_k}}\) is primarily governed by the Gaussian component, whereas the polynomial component contributes negligibly.

\providecommand{\noopsort}[1]{}\providecommand{\singleletter}[1]{#1}%

\end{document}